\documentclass[]{aa}
\usepackage{graphicx}

\begin{document}

\thesaurus{2
              (11.04.1;  
               11.17.1;  
               11.17.4 Q0122$-$380;  
               12.03.3;  
               12.05.1;  
               12.12.1   
              )}

\title{Redshifts of 10 quasar candidates in the field of the rich
       absorption line quasar Q0122$-$380\footnotemark[1]}

\author{R.~A. Jansen \and P. Jakobsen}
\offprints{R.~A. Jansen; {rjansen@astro.estec.}\-{esa.nl}}

\institute{Astrophysics Division, Space Science Department of ESA,
           ESTEC, NL-2200 AG Noordwijk, The Netherlands\\
           email: rjansen@astro.estec.esa.nl, pjakobse@astro.estec.esa.nl
}

\date{Received 28 February 2000 / Accepted 26 April 2000}

\authorrunning{R.~A. Jansen \& P. Jakobsen}
\titlerunning{Quasar candidates in the field of Q0122$-$380}

\maketitle

\renewcommand{\thefootnote}{$\fnsymbol{footnote}$}
\footnotetext[1]{Based on optical spectroscopy obtained at the European
		 Southern Observatory, La Silla, Chile}
\renewcommand{\thefootnote}{$\arabic{footnote}$}

\markboth{R.~A. Jansen \& P. Jakobsen}
	 {Quasar candidates in the field of Q0122$-$380}


\begin{abstract}

We have obtained low resolution ($\sim$20~\AA\ FWHM) slit spectra of 10
quasar candidates located within one degree of the $z=2.181$ quasar
Q0122$-$380 with the objective of searching for signs of large scale
structure matching the intervening rich absorption complexes seen over
the redshift range $1.81 \la z \la 1.97$ toward this object.  Of the 8
confirmed quasars, 4 turn out to have redshifts $z<1.8$, placing them
well in front of the redshift range of interest.  Two of the three
confirmed quasars at redshift $z>1.8$ show no obvious absorption
matching that of Q0122$-$380 at our spectral resolution and
signal-to-noise ratio.  The third object at a redshift of $z=1.868$
displays strong $z_{\rm abs} \sim z_{\rm em}$ absorption systems at $z
\simeq 1.84$ and $z\simeq 1.86$ and a possibly BAL-like trough at
$z\simeq 1.76$. If not intrinsic in nature, the former two systems 
could potentially be related to the absorption seen in Q0122$-$380, 
albeit over a distance of $50\farcm2$ ($D_\perp \simeq 15\,h^{-1}$ Mpc
at $z\simeq 1.9$).


\end{abstract}

  
\section{Introduction}

The existence of large scale structure at high redshift ($z \ga 1$)
provides an important constraint on theories for the formation of
structure and evolution of the Universe.  One approach to probing for
such structure is through the study of intervening metal line absorption
systems in quasar spectra.  Such systems are believed to trace galaxies
through their extended gaseous halos. 

Statistical analysis of the redshift distribution of available samples
of quasar absorption systems suggest that large scale clustering on
comoving scales up to $\sim$100 Mpc may have been in place already at
$z$$\sim$2--3 (e.g.\  Quashnock et al.\  \cite{QVY96}). 

The complementary technique of probing for large scale structure in the
plane of the sky by searching for correlated absorption in adjacent
lines of sight is hampered by the relatively low density of high
redshift quasars bright enough for detailed absorption line work. 
Nonetheless, several potential high redshift 'absorption superclusters'
spanning tens of Mpc on the sky have been identified in this manner. 
These include the $z\simeq1.65$ absorption systems seen toward
PKS~0273$-$233 (Foltz et al.\  \cite{Fetal93}); the two pairs of damped
Ly$\alpha $\ systems seen at $z\simeq 2.38$ and $z\simeq 2.85$ toward
Q2138$-$4427 and Q2139$-$4434 (Francis \& Hewitt \cite{FH93}; Francis
et al.\  \cite{Fetal96}); the apparent structures at $z\simeq 2.3$ and
$z\simeq2.5$ detected in a dense quasar field near the south Galactic
pole by Williger et al.\  (\cite{Wetal96}); and the well-studied case of
the strong absorption spanning $1.8 \la z \la 2.2$ in the field of the
quasar pair Tol~1037$-$2703/1038$-$2712 (Jakobsen et al.\  \cite{Jetal86};
Dinshaw \& Impey \cite{DI96}; Lespine \& Petitjean \cite{LP97}; and
references therein). 

With the aim of searching for further such cases of intervening
high-redshift superclusters Romani et al.\  (\cite{RFS91})
searched the quasar catalogs for suitable background objects near
quasars known to display rich metal line absorption systems.  One of the
most promising fields identified by Romani et al.\  is that of the
$z=2.181$ quasar Q0122$-$380, an object whose absorption spectrum
contains at least seven C\,{\sc iv}\ systems between $1.81 \la z \la 1.98$
(Carswell et al.\  \cite{Cetal82}) and happens to lie within a field in
which Savage et al.\  (\cite{Setal84}) have carried out a deep objective
prism quasar search.  Q0122$-$380 is therefore surrounded by 11 quasar
candidates of brightness $V\simeq$19--20 within a $1^\circ$ radius,
corresponding to comoving distances $D_\perp \la 44\,h^{-1}$ Mpc at
$z\simeq1.95$. 

In this paper we present exploratory slit spectra and redshifts of these
quasar candidates.  As it turns out, several of the objects of interest
are either not confirmed as quasars, or lie at significantly lower
redshift than indicated by their preliminary catalog entries, thereby
rendering the field toward Q0122$-$380 rather less promising for the
purpose of searching for high redshift superclusters than originally
thought.

\begin{table*}[ht!]
\setlength{\tabcolsep}{3pt}
\begin{center}
\parbox[b]{0.725\linewidth}{
   \caption[]{Objects observed in the field of Q0122$-$380}
   \label{Q-Tab.coordsz}
}
\begin{tabular*}{0.725\linewidth}[h]{ p{0.15\linewidth} lll lll r r r r }
\hline\noalign{\smallskip}
\multicolumn{1}{c}{Object$\ \qquad\ $}  &
\multicolumn{3}{c}{$\alpha$(J2000)} &
\multicolumn{3}{c}{$\delta$(J2000)} &
\multicolumn{1}{c}{$\quad m_{\rm V}$} &
\multicolumn{1}{c}{$\quad z^{\ast}{}^{\dag}$} &
\multicolumn{1}{c}{$z_{\rm em}$} &
\multicolumn{1}{c}{$\quad\theta^\ddag$} \\
\noalign{\smallskip}
\hline\noalign{\smallskip}
\ Q0122$-$380\        & 01$^h\!\!$&24$^m\!\!$&17\fs52 &
		     $-$37\degr$\!\!$&44\arcmin$\!\!$&27\farcs0\ \ & $\quad$16.5 & $\quad$2.181 & 2.189$\pm$0.006 & $\quad$ \phantom{0.00} \\
\noalign{\medskip}                                               
\ Q0117$-$379\        & 01 & 19 & 56.47 & $-$37 & 38 & 39.\,3\ \ & $\quad$20.0 & $\quad$1.48~ & 1.484$\pm$0.007 & $\quad$52.0 \\
\ Q0117$-$380\        & 01 & 19 & 45.65 & $-$37 & 48 & 27.\,1\ \ & $\quad$18.9 & $\quad$2.02~ & 2.020$\pm$0.007 & $\quad$53.9 \\
\ Q0118$-$377\        & 01 & 20 & 43.15 & $-$37 & 28 & 38.\,5\ \ & $\quad$19.0 & $\quad$0.34~ & 1.728$\pm$0.007 & $\quad$45.3 \\
\ Q0120$-$3781\       & 01 & 22 & 35.50 & $-$37 & 32 & 56.\,9\ \ & $\quad$19.5 & $\quad$2.15~ & 2.124$\pm$0.007 & $\quad$23.2 \\
\ Q0120$-$3785\       & 01 & 22 & 45.05 & $-$37 & 35 & 31.\,6\ \ & $\quad$19.5 & $\quad$2.17~ & 1.516$\pm$0.004 & $\quad$20.3 \\
\ Q0121$-$373\        & 01 & 24 & 15.06 & $-$37 & 05 & 13.\,9\ \ & $\quad$19.4 & $\quad$1.49~ & 1.055$\pm$0.009 & $\quad$39.2 \\
\ Q0124$-$373\        & 01 & 26 & 15.52 & $-$37 & 07 & 46.\,1\ \ & $\quad$19.8 & $\quad$0.88~ & 0.920$\pm$0.010 & $\quad$43.5 \\
\ Q0125$-$376\        & 01 & 28 & 05.78 & $-$37 & 22 & 37.\,6\ \ & $\quad$19.0 & $\quad$1.84~ & 1.868$\pm$0.008 & $\quad$50.2 \\
\noalign{\medskip}
\ \   ~0121$-$379$^a$ & 01 & 23 & 25.89 & $-$37 & 42 & 23.\,6\ \ & $\quad$19.7 & $\quad$2.21~ & \multicolumn{1}{c}{$\cdots$} & $\quad$10.4 \\
\ \   ~0123$-$372$^a$ & 01 & 25 & 27.10 & $-$36 & 58 & 06.\,1\ \ & $\quad$20.3 & $\quad$2.13~ & \multicolumn{1}{c}{$\cdots$} & $\quad$48.4 \\
\ \   ~0117$-$378$^b$ & 01 & 19 & 56.82 & $-$37 & 37 & 28.\,4\ \ & $\quad$20.0 & $\quad$2.25~ & 
                                                             \multicolumn{1}{c}{$\cdots$} & \multicolumn{1}{c}{$\quad\cdots$} \\
\noalign{\smallskip}\hline\noalign{\smallskip}
\end{tabular*}
\parbox[b]{0.725\linewidth}{
   \noindent
   $^\dag$ Preliminary redshift listed in Hewitt \& Burbidge (1993)\\
   $^\ddag$ Angular distance from Q0122$-$380 in arcmin.\\
   $^a$ Object is a star, not a quasar.\\
   $^b$ No object was found at the catalog coordinates.\\
}
\end{center}
\end{table*}
%


\section{Observations}

Table~1 lists the coordinates, magnitudes and preliminary redshifts
$z^\ast$ of all objects listed in the Hewitt \& Burbidge (1993) catalog
and located within a circle of $1^\circ$ radius centered on Q0122$-$380. 
These 11 objects along with Q0122$-$380 itself were observed with the
ESO 3.6~m telescope and ESO Faint Object Camera and Spectrograph
(EFOSC1) on 1995 September 28.  EFOSC1 was equipped with a thinned
back-side illuminated TEK CCD with 512$\times $512, 27$\mu$m pixels.  We
used a 230~\AA\, mm$^{-1}$ (B300) grism in combination with a 2\arcsec\
wide slit to obtain spectra covering the wavelength range
3750--6950~\AA\ at $\sim$20~\AA\ (FWHM) resolution.  The spatial
resolution is 0.61\arcsec\ pixel$^{-1}$.

\begin{figure*}[ht!]
\centerline{
   \hfill\includegraphics[width=\textwidth]{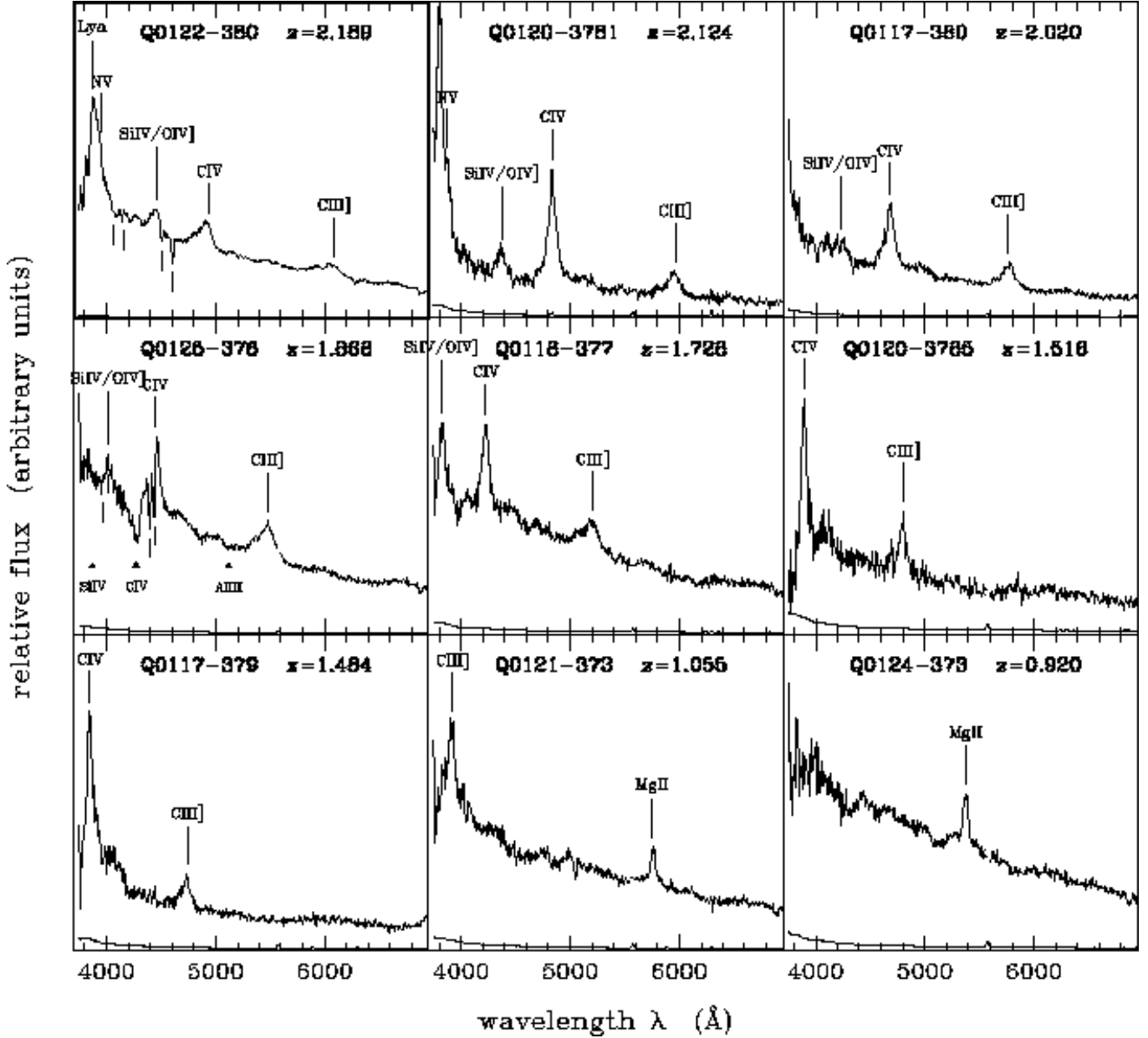}\hfill
}\par
\parbox[t]{\textwidth}{
   \caption[]{Spectra of the eight confirmed quasars. For reference the
	spectrum of Q0122$-$380 is given in the top-left panel. The
	strongest emission features and the corresponding emission-line
	redshifts are indicated. The spectra span the range
	3750--6950~\AA\ and the resolution is $\sim$20~\AA. The fluxes
	have been normalized to the flux in a 100~\AA\ region centered
	on 5470~\AA.  Below each spectrum, the 1-sigma noise level is
        shown as a thin curve. The two C\,{\sc iv}\ and Si\,{\sc iv}\
	absorption complexes seen near $z\simeq 1.91$ and $z \simeq
	1.97$ toward Q0122$-$380 are marked in its spectrum. In the
	spectrum of Q0125$-$376 the $z_{\rm abs} \sim z_{\rm em}$
	absorption systems at $z\simeq 1.84$ and $z\simeq 1.86$ are
	indicated by vertical bars, the BAL-like absorption troughs by
	small triangles.
   }
}
\end{figure*}
%

The seeing conditions on both nights were poor.  Nevertheless, the high
throughput of EFOSC1 ensured good signal-to-noise (S/N) spectra of
objects as faint as $m_{\rm V}=20$ (S/N$\geq$8 in the continuum near
5500~\AA).  For most objects we took a single 1200~s exposure.  For
0117$-$379, being the faintest target at $m_{\rm V}=20.0$, we exposed for
1800~s.  Prior to each spectroscopic observation, EFOSC1 was used in
direct (filterless) imaging mode for target verification and automated
slit acquisition.  One quasar candidate, 0117$-$378, was not found at or
near its catalog coordinates. 

The data were reduced within the IRAF environment, following standard
techniques.  The spectra were put on a relative flux scale based on the
standard stars LDS~235/EG~63 and LTT~2415 (Baldwin \& Stone 1984; Stone
\& Baldwin 1983).  The absolute calibration is ill determined, due to
variations in seeing and atmospheric extinction over the night. 

The resulting spectra of the 8 confirmed quasars are shown in Fig.~1
along with the relevant emission line identifications.  The measured
redshifts given in Table~1 were determined by averaging the redshifts
measured for individual emission lines (e.g.\ Ly$\alpha $,
N\,{\sc v}~$\lambda$1240, C\,{\sc ii}~$\lambda$1335, 
Si\,{\sc iv}/O\,{\sc iv}]~$\lambda$1400, C\,{\sc iv}~$\lambda$1549, 
He\,{\sc ii}~$\lambda$1640, Al\,{\sc iii}~$\lambda$1857, 
C\,{\sc iii}]~$\lambda$1909, and Mg\,{\sc ii}~$\lambda$2799).  For the
wavelength of an emission line we adopted the average of the wavelengths
of the peak (maximum signal) and of the center of a gaussian fit to the
line, both determined after subtraction of the quasar continuum. 

Two of the quasar candidates, 0121$-$379 and 0123$-$372, turn out to be
Galactic stars.  Their spectra are shown in Fig.~2.  Unfortunately, of
the observed quasar candidates 0121$-$379 would have been the quasar
closest to Q0122$-$380, at an angular separation of $10\farcm4$.


\section{Discussion}

Romani et al.\ (1991) originally drew attention to the rich absorption
line quasar Q0122$-$380 on the basis that it was surrounded by a number
of quasar candidates closer than $1^\circ$.  Of the 11 objects listed in
Table~1, seven had preliminary redshifts $z>1.8$, placing them near or
behind the absorption seen between $1.81 \la z \la 1.98$ toward
Q0122$-$380. 

Of these seven high redshift candidates, only three (Q0117$-$380,
Q0120$-$3781 and 0125$-$376) are confirmed as $z>1.8$ quasars.  Of the
remainder, one (0117$-$378) could not be located, two (0121$-$379 and
0123$-$372) are identified as stars, and another (Q0120$-$3785) turns
out to be at a lower redshift of $z=1.52$.  The four quasar candidates
with lower preliminary redshifts $z<1.8$ are all confirmed as such.

\begin{figure}[ht!]
\centerline{
   \hfill\includegraphics[width=0.375\textwidth]{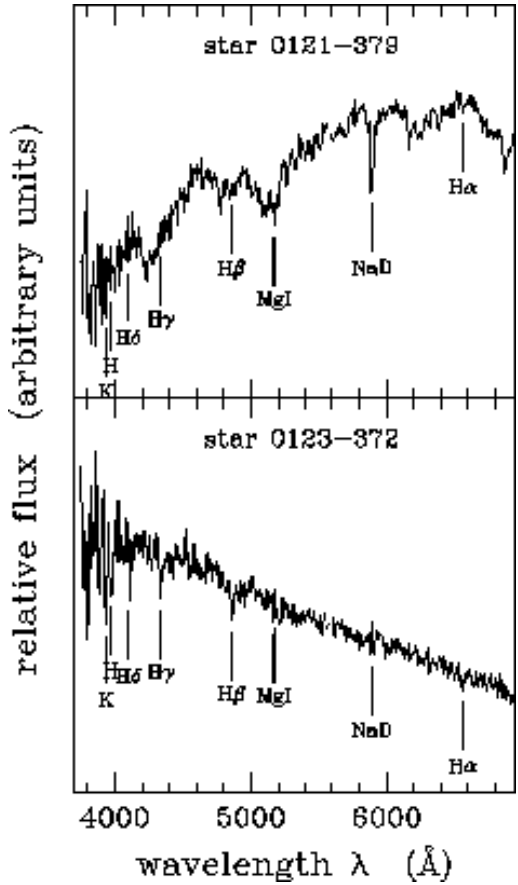}\hfill
}\par
\parbox[t]{0.485\textwidth}{
   \caption[]{Spectra of the two stars misclassified as quasars in
	Hewitt \& Burbidge (1993).  The stellar absorption features are
	indicated.
   }
}
\end{figure}
%

As is evident from Fig.~1, our low resolution spectra of two of the
three confirmed $z>1.8$ quasars (Q0120$-$3781 and Q0117$-$380) reveal no
obvious C\,{\sc iv}\ absorption features in the wavelength range 4350-4600~\AA\
that could potentially be associated with the absorption spanning $1.81
\la z \la 1.97$ toward Q0122$-$380.  However, the line detection limit
of our spectra is only $W_\lambda \ga 8$~\AA\ at these wavelengths. 

A more promising case is that of the final object, Q0125$-$376, whose
redshift of $z=1.868$ lies close to that of the absorption complex at $z
\simeq 1.91$ in Q0122$-$380.  Moreover, our spectrum of Q0125$-$376
(which has a better S/N ratio than those of Q0120$-$3781 and
Q0117$-$380) shows three strong ($W_\lambda \ga 9$~\AA) absorption
features at $\lambda\simeq 4395$~\AA, $\lambda\simeq 4435$~\AA\ and
$\lambda\simeq 4275$~\AA, respectively.  The former two features are
almost certainly due to C\,{\sc iv}\ absorption from two $z_{\rm abs}
\sim z_{\rm em}$ systems at $z\simeq 1.837$ and $z\simeq 1.864$, an
interpretation that is further strengthened by plausible detections of
matching Si\,{\sc iv}\ lines at shorter wavelengths.  Based on a matching
weaker feature seen at the anticipated position of Al\,{\sc iii}, we
tentatively identify the third feature as the C\,{\sc iv}\ trough of a
weak and possibly detached BAL-like complex at $z\simeq 1.76$\/. 

\newpage

The two $z_{\rm abs} \sim z_{\rm em}$ systems seen in Q0125$-$376 could
conceivably be associated with the absorption seen
toward Q0122$-$380, falling squarely between the $z \simeq 1.91$ complex
and the weaker $z \simeq 1.814$ system detected in that object by
Carswell et al.\ (1982).  In a standard cosmological model with
$q_0=0.5$ the angular distance between Q0125$-$376 and Q0122$-$380 of
$50\farcm2$ at $z=1.9$, corresponds to a projected comoving separation
of D$_\perp = 36\,h^{-1}$ Mpc $(\mbox{H$_0$}=h\,100\,\mbox{km s$^{-1}$
Mpc$^{-1}$})$, which is comparable to the extent of local superclusters
of galaxies. 

On the other hand, the presence of possibly BAL-like absorption at
lower redshift would argue that the two $z_{\rm abs} \sim z_{\rm em}$
systems seen in Q0125$-$376 are intrinsic in nature.

While higher resolution observations would be required to further
delineate these possibilities and properly map the absorption toward   
the three confirmed $z>1.8$ quasars above, the exploratory
observations presented here already make it clear that the field
surrounding Q0122$-$380 is not as promising for searching for large
scale structure at high redshift as had initially been hoped.


%
\end{document}